\documentclass[aps,prl,twocolumn,superscriptaddress]{revtex4}%
\usepackage{amsmath}
\usepackage{graphicx}%
\usepackage{amsfonts}%
\usepackage{amssymb}

\begin{document}

\title{Universal features of charge and spin order in a half-doped layered perovskite.
 }

\author{I. A. Zaliznyak}
\affiliation{Department of Physics, Brookhaven National Laboratory, Upton,
 New York 11973-5000}

\author{J. M. Tranquada}
\affiliation{Department of Physics, Brookhaven National Laboratory, Upton,
 New York 11973-5000}

\author{G. Gu}
\affiliation{Department of Physics, Brookhaven National Laboratory, Upton,
 New York 11973-5000}

\author{R. W. Erwin}
\affiliation{National Institute of Standards and Technology, Gaithersburg,
 Maryland 20899}

\author{Y. Moritomo}
\affiliation{Department of Applied Physics, Nagoya University, Nagoya 464-8603,
Japan}


\begin{abstract}

We have investigated the peculiar structure of charge and spin ordering in the
half-doped layered perovskite oxide La$_{1.5}$Sr$_{0.5}$CoO$_4$ by elastic
neutron scattering. Two samples with different preparation histories were
studied. We find that the generic features of the ordered states, such as their
short-range, glassy nature and the spin incommensurability, are
sample-independent. At the same time, some subtle features of the ordered
phases, such as the correlation lengths, differ.

\end{abstract}


\maketitle


Recently much attention has been paid to exploring the charge/orbital- and
spin-ordered phases in the charge-doped strongly-correlated transition-metal
oxides, such as the superconducting cuprates and the magnetoresistive manganites
(see Refs. \onlinecite{Orenstein2000,Tokura2000} for a review). Competition of
the spin, orbital and/or the charge/valence degrees of freedom results in
non-trivial ground states, and leads to many fascinating physical properties,
such as a magnetic-field-driven metal-insulator transition, etc. At small
doping, the charges often segregate into lines, forming a stripe-like
superstructure \cite{TranquadaWakimotoLee}, which may be important for the
high-temperature superconductivity in cuprates. At half-doping, $x=0.5$, there
is a natural opportunity for a robust valence-ordered state with the
checkerboard filling of the two-dimensional (2D) square MO$_2$ lattice (M = Cu,
Mn, Ni, Co, ... is a 3d metal) by M$^{2+}$/M$^{3+}$ or M$^{4+}$/M$^{3+}$ ions.
At high doping, close to $x \sim 0.5$, such a phase with the checkerboard charge
order (CO) is often stabilized by the Jahn-Teller distortion, and is in
competition with the metallic state where the charges are delocalized. Such a
competition is at the heart of the strong response (CMR) of the doped
pseudocubic manganites to a magnetic field, which favors the metallic phase by
inducing ferromagnetism, \cite{Schiffer1995}.

The CO and the spin-ordered (SO) phases were recently characterized in a number
of the half-doped materials, in particular, in the layered perovskites
La$_{0.5}$Sr$_{1.5}$MnO$_4$ \cite{Sternlieb1996}, La$_{1.5}$Sr$_{0.5}$CoO$_4$
\cite{Zaliznyak2000,Zaliznyak2001}, and La$_{1.5}$Sr$_{0.5}$NiO$_4$
\cite{Kajimoto2003}. All three materials have the tetragonal crystal structure
of the ``HTT'' phase (space group $I4/mmm$, see Fig.~1(a) in Ref.
\onlinecite{Zaliznyak2001}), and with lattice spacings that only differ by
$\lesssim 1\%$. Although an ``orthorhombic'' indexing based on the space group
$F4/mmm$, with a twice larger unit cell, $\sqrt{2}a \times \sqrt{2}a \times c$,
is often employed to account for breaking of the translational symmetry in the
$a-b$ plane by the checkerboard CO superstructure, here we use the $I4/mmm$
notation corresponding to the true crystal lattice.

With decreasing temperature, all three compounds first undergo a transition to
the checkerboard charge-ordered state. It is accompanied by a dramatic increase
of the activation energy in the $T$-dependence of the electrical conductivity,
which is a clear evidence of the charge localization. An important generic
feature of the CO in the layered materials at half-doping is its finite
correlation range. It results from the combination of the random electrostatic
potential introduced by the quenched disorder of the dopant ions, and the
quasi-two-dimensional nature of the elastic couplings between the distortions in
different layers \cite{Zachar2003}. If the coupling between the distortions is
isotropic, as it is in the pseudo-cubic manganites, the CO may become a truly
long-range superstructure.

In all three materials the spin ordering (SO) follows the charge order, but at a
significantly lower temperature. In the manganite, La$_{0.5}$Sr$_{1.5}$MnO$_4$,
the SO is commensurate, with the propagation vector $Q=(1/4,1/4,1)$
\cite{Sternlieb1996}, as for the simple two-sublattice antiferromagnetism on
each of the two fields of the CO checkerboard. In La$_{1.5}$Sr$_{0.5}$CoO$_4$,
which is the subject of this study, the SO was found to be incommensurate, with
the propagation vector $Q=(\zeta,\zeta,1)$ with $\zeta \approx 0.258$
\cite{Zaliznyak2000}. The situation is even more complex and intriguing in the
nickelate. There the SO is also incommensurate, but with $\zeta \approx 0.22$,
smaller than 1/4, and is preceded by the transformation of the CO from a
checkerboard to a stripe-like phase with the propagation vector
$Q=(2\zeta,2\zeta,0)$ \cite{Kajimoto2003}.

The correlation lengths of spin and charge order measured in the above materials
are summarized in Table \ref{table1}. Both SO and CO are much better correlated
in the manganite, where the SO is commensurate. Although a recent X-ray study
\cite{Wakabayashi2001} determined that $\xi_{ab}^{(CO)} = 200(30)$\AA\ and
$\xi_{c}^{(CO)} = 30(2)$\AA\ in La$_{0.5}$Sr$_{1.5}$MnO$_4$, these values are
still about an order of magnitude larger than the CO correlation lengths for the
other two compounds in Table \ref{table1}. The situation with the SO correlation
lengths is similar. Whether the spin incommensurability is in some way related
with this difference in the correlation lengths, though, is not really clear.
Furthermore, a question arises: is there any connection at all between the CO
and SO correlation ranges for a given material?

The difference in the CO correlation lengths measured in different materials can
readily be understood from the generic cross-over phase diagram based on the
random field Ising model (RFIM) \cite{Zachar2003}. It simply indicates that the
effective random field and/or the inter-plane coupling in the quasi-2D
anisotropic RFIM, which is the effective low-energy theory describing the CO
transition, differ. However, because the source of the random field is the
crystallographic disorder, it may also vary from one sample to another of the
same material, with the same nominal stoichiometry. How strong is this effect,
and how does it compare to the variation of $\xi^{(CO)}$, $\xi^{(SO)}$ between
the different materials?

\begin{table}[pt]
\vspace{-0.1in}%
\caption{Correlation lengths (in \AA) of charge and spin order in three
half-doped isostructural layered perovskite oxides.}%
\begin{tabular}{lccccr}
\hline\hline%
 Material & $\xi_{ab}^{(CO)}$ & $\xi_{c}^{(CO)}$ & $\xi_{ab}^{(SO)}$ & $\xi_{c}^{(SO)}$ & Ref. \\
\tableline
 La$_{0.5}$Sr$_{1.5}$MnO$_4$ & $\gtrsim 300$ & $\approx 50$ & $\gtrsim 300$ & $\approx 33$ & \cite{Sternlieb1996} \\
 La$_{1.5}$Sr$_{0.5}$NiO$_4$ & 30(10) & 2(1) & $\approx 120$ & $\approx 13$ & \cite{Kajimoto2003} \\
 La$_{1.5}$Sr$_{0.5}$CoO$_4$ & 23(3) & 8.3(6) & 79(3) & 10.7(3) & \cite{Zaliznyak2000,Zaliznyak2001} \\
\hline\hline%
\end{tabular}
\label{table1}%
\vspace{-0.25in}%
\end{table}


To address these questions, we have studied two samples of the half-doped
cobaltate, La$_{1.5}$Sr$_{0.5}$CoO$_4$, of different size and with different
preparation histories, using elastic neutron scattering. Sample \#1 is a
cylinder of $m_1 \approx 0.5$~g and $\approx 5$ mm long, prepared at Nagoya
University as described in Ref. \onlinecite{Moritomo1997}, and was previously
studied in Refs. \onlinecite{Zaliznyak2000,Zaliznyak2001}. A larger sample \#2
is $\approx 45$ mm long, $m_2 \approx 11$~g cylinder, and was grown in air
atmosphere in the floating zone furnace at Brookhaven National Laboratory. Both
samples were a high quality single crystals with mosaic spread of less than
$0.25^\circ$. Within the accuracy of our measurement they both have the same low
$T$ lattice constants, $a=3.830(3)$\AA\ and $c=12.51(2)$\AA.

\begin{figure}[pbb]
\label{fig1}
\vspace{-0.1in}%
\begin{center}
\includegraphics[height=2.5in,width=3.5in]{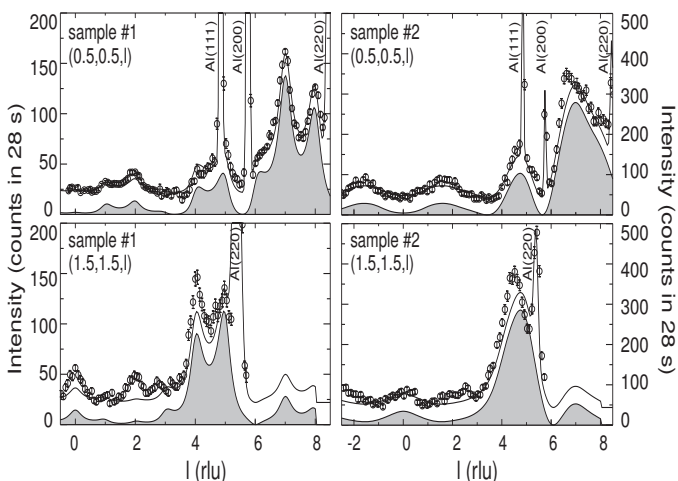}%
\end{center}
\vspace{-0.3in}%
\caption{Dependence of the structural elastic diffuse scattering accompanying
the charge order on the wavevector transfer along the $c$-axis in two samples.
Al(111), Al(200) and Al(220) mark the corresponding peaks arising from the
``parasitic'' scattering by the aluminium in the sample environment.}
\end{figure}

Measurements of the wave-vector dependence of the elastic scattering associated
with charge and spin ordering were performed on the BT2 spectrometer at NIST
Center for Neutron Research. The incident and the scattered neutron energy, $E_i
= E_f = 14.7$ meV, was defined by PG(002) reflections, with two PG filters
suppressing the higher-order contamination. Beam collimations were $\approx 60'$
and $\approx 100'$ before the monochromator and after the analyzer,
respectively, and $20'-20'$ around the sample. The sample was mounted in a
displex refrigerator with its $(hhl)$ reciprocal lattice zone in the horizontal
scattering plane, and kept at $T = 10(2)$ K. To establish the similarity of the
temperature dependence of the SO scattering in the two samples, some additional
measurements were performed using the SPINS cold-neutron spectrometer.


\begin{figure}[ptb]
\label{fig2}
\begin{center}
\includegraphics[width=3.2in,height=3.5in]{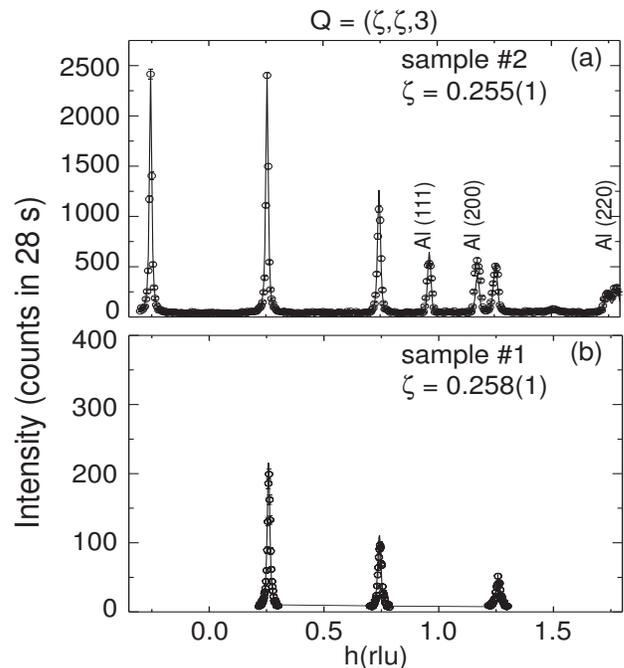}
\end{center}
\vspace{-0.3in}%
\caption{Incommensurate magnetic peaks arising from the short-range spin order
in two samples, (a) \# 1, and (b) \#2. Temperature dependence of the magnetic
intensity is essentially identical in both cases, with an onset at $T_s \approx
30$ K.}
\vspace{-0.25in}%
\end{figure}

The dependence of the CO scattering on the wave vector transfer along the
$c$-axis for the two samples is shown in Fig. 1 (left panel, sample \# 1; right
panel, \#2). The scans through the CO peaks at $Q=(h+0.5,h+0.5,l)$ in the
$ab$-plane, differ essentially only in their intensity, and are not shown. The
CO diffuse scattering was analyzed following the procedure described in the Ref.
\onlinecite{Zaliznyak2001}. The solid curves in Figure 1 show the corresponding
resolution-corrected fits to the ``lattice Lorentzian" scattering function which
describes the diffuse scattering by a short-range superstructure (shaded areas
show the calculated intensity in the absence of the background).

It is clear from Figure 1 that the long-period $l$-dependence of the diffuse
intensity is similar in both samples. This indicates that in both cases it
results from the same kind of local distortion of the CoO$_6$ octahedra. In
fact, the fit shows that the corresponding displacements of the oxygen and Sr/La
ions are, within its accuracy, the same in both samples,
$\varepsilon^{x,y}_{O(1)} = 0.044(4)$ \AA, $\varepsilon^{z}_{O(2)} =
0.080(8)$\AA, and $\varepsilon^{z}_{La/Sr} = 0.010(2)$\AA\ (cf. Refs.
\onlinecite{Zaliznyak2000,Zaliznyak2001}; here we have allowed for a small
modulation of the La/Sr position which slightly improves the fit). What makes
the scattering patterns from the two samples look quite distinct, is the
difference in the CO correlation range between the $ab$ planes,
$\xi_{c}^{(CO)}$.

The correlation lengths for the charge [$\xi_{ab}^{(CO)}, \xi_{c}^{(CO)}$] and
the spin order [$\xi_{ab}^{(SO)}, \xi_{c}^{(SO)}$] refined from our measurements
are summarized in Table \ref{table2}. The inter-plane CO correlation length in
the sample \#1  is somewhat larger than the distance between the neighboring
$ab$ planes, $c/2$, and the CO peaks at integer $l$ are well separated. On the
other hand, in the case of sample \# 2 there are no such well-defined peaks.
This means that the inter-plane correlations are not developed at all, i.e. the
correlation length is much smaller than the distance between the neighboring
$ab$ planes. In our refinement we fixed $\xi_{c}^{(CO)} = 2.5$\AA. At the same
time, the CO correlation range within the $ab$ layers is essentially the same
for both samples. This agrees with the expectation from the anisotropic RFIM
\cite{Zachar2003}. In the strongly anisotropic regime,
$\xi_{c}^{(CO)}/\xi_{ab}^{(CO)} \ll 1$, the in-plane correlation lenghth,
$\xi_{ab}^{(CO)}$, is essentially determined by the 2D RFIM physics, and is not
sensitive to the developing inter-plane correlations.

Typical scans characterizing the dependence of the magnetic scattering
associated with the spin order on the wave-vector transfer in the $ab$ plane are
shown in the Figure 2. The solid lines are the appropriate resolution-corrected
``lattice Lorentzian'' fits, as in Ref. \onlinecite{Zaliznyak2000}. The spin
structure is described as a damped flat spiral polarized in the $ab$ plane, with
the propagation vector ${Q} = (\zeta, \zeta, 1)$. The $l$-scans (not shown) have
much broader peaks centered at odd $l$ (the inter-plane SO correlation length,
$\xi_{c}^{(SO)}$, is also much shorter than that in the plane,
$\xi_{ab}^{(SO)}$). They are also well described by the model.

Although the charge order is somewhat better correlated in sample \# 1, the
refinement shows that the spin order correlation lengths are appreciably, by
more than a factor of 1.5, larger in sample \#2. This supports the conclusion
that, although the CO correlation range in the $ab$ plane is quite short, the
resulting disorder is still far from the 2D percolation threshold for the spin
system. Therefore, it is not the source of the SO finite correlation length
which, most probably, originates from the frustrated spin interactions
\cite{Zaliznyak2000}. This conclusion is further corroborated by the observation
that the SO incommensurability $\zeta$ appears smaller in the sample \# 2, where
the spin correlations are appreciably stronger (although the difference is
small, it is well beyond the most conservative error bars which are shown in
Table \ref{table2}).

\begin{table}[pt]
\vspace{-0.1in}%
\caption{Charge and spin order correlation lengths (\AA) and the spin
incommensurability $\zeta$ (lattice units) observed in two different
La$_{1.5}$Sr$_{0.5}$CoO$_4$ single crystal samples.}%
\begin{tabular}{lccccc}
\hline\hline%
 Sample & $\xi_{ab}^{(CO)}$ & $\xi_{c}^{(CO)}$ & $\xi_{ab}^{(SO)}$ & $\xi_{c}^{(SO)}$ & $\zeta$ \\
\tableline
 \# 1 (small) & 19(2) & 8(1) & 79(3) & 9(1) & 0.258(1) \\
 \# 2 (large) & 18(1) & $\lesssim 2.5$ (fixed) & 125(6) & 15.5(5) & 0.255(1) \\
\hline\hline%
\end{tabular}
\label{table2}%
\vspace{-0.25in}%
\end{table}


To summarize, we find that the generic, qualitative features of the charge and
spin ordering in the half-doped layered perovskite La$_{1.5}$Sr$_{0.5}$CoO$_4$,
such as their short-range, glassy nature and the spin incommensurability are
common for the two samples that we have studied. In fact, they are also shared
by the other layered half-doped materials (Table \ref{table1}). However, we also
find some appreciable differences between our two samples with different
preparation histories. The most dramatic is the difference in the correlation
lengths. While the in-plane CO correlations are essentially the same, the CO
correlation between the layers is significantly weaker in the sample \# 2. This
can be understood on the basis of the anisotropic RFIM \cite{Zachar2003} that
describes the charge ordering in the half-doped materials. It implies that the
weak residual inter-layer CO coupling is sensitive to the impurities and the
imperfections in the crystal and is sample dependent.

For the spin order, the correlation lengths are as different between the two
samples, as those between the sample \#1 and the nickelate,
La$_{1.5}$Sr$_{0.5}$NiO$_4$. In fact, the correlation lengths of the
checkerboard charge order and of the SO in the nickelate are essentially the
same as in our sample \# 2 of La$_{1.5}$Sr$_{0.5}$CoO$_4$. The spin
incommensurability, though, is completely different, and the charge-stripe phase
is not present in either of the cobaltate samples. Finally, the smaller
incommensurability $\zeta$ in the sample \#2 with stronger spin correlations
seems to indicate that these are related, and, perhaps, $\zeta$ vanishes if the
SO is truly long range. This may explain, why the spin structure is apparently
commensurate in the manganite, La$_{0.5}$Sr$_{1.5}$MnO$_4$, where the SO
correlation length is at least twice larger than in our sample \#2.


We thank O. Zachar for valuable discussions, and acknowledge the support under
the Contract DE-AC02-98CH10886, Division of Materials Sciences, US Department of
Energy. The work on SPINS was supported by NSF through DMR-9986442.

\end{document}